\documentclass[prl,twocolumn,showpacs,amsmath,amssymb]{revtex4}
\usepackage{graphicx}
\usepackage{dcolumn}
\usepackage{bm}

\usepackage{color}

\usepackage{ulem}

\begin{document}

\title{Proton-neutron pairing vibrations in $N=Z$ nuclei: 
Precursory soft mode of isoscalar pairing condensation
}

\author{Kenichi Yoshida}
\affiliation{Graduate School of Science and Technology, Niigata University, Niigata 950-0913, Japan
}%

\date{\today}

\begin{abstract}
$L=0$ proton-neutron ($pn$) pair-addition and pair-removal strengths 
in $^{40}$Ca and $^{56}$Ni are investigated by means of the 
$pn$ particle-particle random-phase approximation employing a Skyrme energy-density functional. 
It is found that the collectivity of the lowest $J^\pi = 1^+$ state in the adjacent odd-odd nuclei  
becomes stronger as the strength of the isoscalar ($T=0$) pairing interaction increases. 
The results suggest the emergence of the $T=0$ $pn$-pairing vibrational mode 
as a possible critical phenomenon toward the $T=0$ pairing condensation. 
\end{abstract}

\maketitle

The pairing correlation plays a central role in low-energy nuclear phenomena; 
such as the ground state spin, 
staggering in the systematics of the binding energies, 
the low-lying quadrupole collective dynamics, and the spontaneous fission half-lives~\cite{ber12}. 
The correlation is so strong that 
the fluctuations of the pairing gap around its zero equilibrium value develop in 
nuclei near the closed shell, and 
the systems get deformed eventually in gauge space when more nucleons are added~\cite{bro05}. 
The collective pairing vibration emerging in the closed-shell nuclei 
is thus associated with the occurrence of the pairing condensation.

It is in the isovector and spin-singlet ($T=1, S=0$) channel 
that the pairing correlation has been extensively studied. 
With the advent of the radioactive-isotope beam technology, 
the heavy proton-rich nuclei along the $N=Z$ line have received considerable attention. 
Of particular interests are location of the proton drip line and the extra binding mechanism 
called the Wigner energy~\cite{lun03}. 
The isoscalar and spin-triplet ($T=0, S=1$) pairing correlation 
is expected to be visible in $N \sim Z$ nuclei 
because the shell structures around the Fermi levels of both neutrons and protons are similar to each other 
and the spatial overlap between the neutron and proton single-particle wave functions would be large 
to form a proton-neutron ($pn$) Cooper pair~\cite{goo98}.   
As a consequence of the strongly attractive $pn$ interaction in the $^3S_1$ channel, 
the possible $T=0$ pairing condensate has been discussed 
in heavy $N \sim Z$ nuclei theoretically~\cite{sat97,goo99,ber10,gez11}.  

The experimental fingerprint of the $T=0$ pairing condensation has been under debate 
though there have been numerous experimental attempts~\cite{fra14}. 
This is because the spin-orbit splitting suppresses to couple a spin-triplet pair 
in the ground state~\cite{pov98}.
In Ref.~\cite{mac00}, Macchiavelli {\it et al}. tried to extract the experimental excitation energy 
and the collectivity of the $T=0$ and $T=1$ $pn$-pair excited states. 
Their analysis is based on the subtraction of 
average properties including the symmetry energy and 
comparison to the single-particle level spacing. 
The Hamiltonian employed for describing the $pn$ pair excitations contains 
the schematic separable interactions of the $T=0$ and $T=1$ types with two levels. 
Then, they found any appreciable collectivity in the $T=0$ channel unlikely in $^{56}$Ni.

The interplay between the $T=0$ and $T=1$ pairing correlations 
in the $pn$-pair transfer strengths has been investigated by employing a solvable model~\cite{dus86}.
I investigate in the present article the possibility 
of a collective $T=0$ $pn$-pairing vibrational mode in the ``normal" phase where the $T=0$ pairing gaps are zero.
The $pn$ pair excitations are described microscopically 
based on the nuclear energy-density functional (EDF) method, where the 
global properties and the shell effects are taken into account on the same footing. 
More precisely, the $pn$-pairing vibrational modes are obtained out of the solutions of 
the $pn$ particle-particle random-phase approximation (ppRPA) equation, 
and are described as elementary modes of excitation generated by 
two-body interactions acting between a proton and a neutron. 
Then, I show that the strongly collective $T=0$ $pn$-pairing vibrational mode emerges 
when the interaction is switched on.

In a framework of the nuclear EDF method employed, 
the $pn$-pair-addition vibrational modes are described as 
$|Z+1, N+1; \lambda \rangle = \hat{\Gamma}^\dagger_\lambda |Z,N \rangle$ 
with the RPA phonon operator $\hat{\Gamma}^\dagger_\lambda  = \sum_{mn}
X_{\lambda, mn} \hat{a}^\dagger_{\pi,m} \hat{a}^\dagger_{\nu,n} 
-  \sum_{ij} Y_{\lambda,ij} \hat{a}^\dagger_{\nu,j} \hat{a}^\dagger_{\pi,i} $.  
Here $a^\dagger_{\pi,m} (a^\dagger_{\nu,n})$ 
create a proton (a neutron) in the single-particle level $m$ $(n)$ above the Fermi level, 
and 
 $a^\dagger_{\pi,i} (a^\dagger_{\nu,j})$ create a  proton (a neutron) in the single-particle level $i $ $(j)$ below the Fermi level. 
The first and second terms correspond to the particle-particle (pp) and hole-hole (hh) excitations, respectively. 
Greek indices $\alpha, \beta$ will be used for indicating the particle and hole states collectively.
The single-particle basis is obtained as 
a self-consistent solution of the Skyrme-Hartree-Fock (SHF) equation. 
The SHF equation is solved in cylindrical coordinates
$\boldsymbol{r}=(r,z,\phi)$ with a mesh size of
$\Delta r=\Delta z=0.6$ fm and with a box
boundary condition at $(r_{\mathrm{max}},z_{\mathrm{max}})=(14.7, 14.4)$ fm.
The axial and reflection symmetries are assumed in the ground state. 
More details of the calculation scheme are given in Ref.~\cite{yos13}.

In the present calculation, the SGII interaction is used for the particle-hole (ph) channel because 
the spin-isospin properties were considered to fix the coupling constants 
entering in the EDF~\cite{gia81}. 
For the pp channel, the density-dependent contact interactions are employed: 
\begin{align}
& v_{\rm{pp}}^{T=0}(\boldsymbol{r} \sigma \tau, \boldsymbol{r}^\prime \sigma^\prime \tau^\prime) \notag \\
&= f \times V_0 \dfrac{1+P_{\sigma}}{2}\dfrac{1-P_{\tau}}{2} \left[ 1- \dfrac{\rho(\boldsymbol{r})}{\rho_0}\right] 
\delta(\boldsymbol{r}-\boldsymbol{r}^\prime), \label{v_pp_0}\\
& v_{\rm{pp}}^{T=1} (\boldsymbol{r}\sigma \tau, \boldsymbol{r}^\prime \sigma^\prime \tau^\prime ) \notag \\ 
&= V_0 \dfrac{1-P_{\sigma}}{2}\dfrac{1+P_{\tau}}{2} \left[ 1- \dfrac{\rho(\boldsymbol{r})}{\rho_0}\right] 
\delta(\boldsymbol{r}-\boldsymbol{r}^\prime), \label{v_pp_1}
\end{align}
where $\rho_0 = 0.16$ fm$^{-3}$ and $\rho(\boldsymbol{r})=\rho_\nu(\boldsymbol{r})+\rho_\pi(\boldsymbol{r})$. 
The pairing strength is fixed as $V_0 = -390$ MeV fm$^3$. 
A procedure to determine $V_0$ will be explained below.
The factor $f$ appearing in  Eq.~(\ref{v_pp_0}) is changed to see an effect of the interaction in 
the $T=0$ channel~\cite{bai13}.

\begin{figure}[t]
\begin{center}
\includegraphics[scale=0.26]{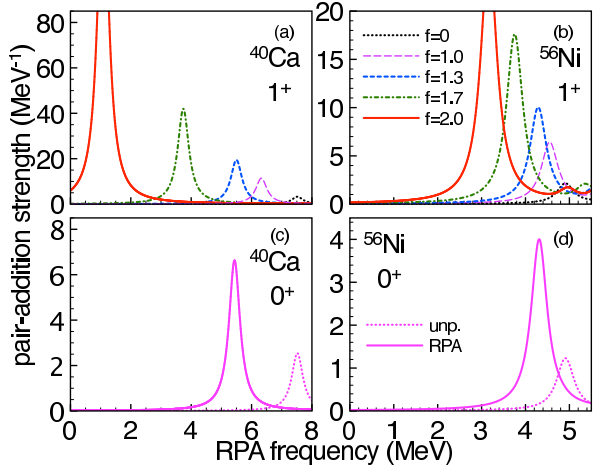}
\caption{(Color online) $pn$ pair-addition strengths of $^{40}$Ca $\to$ $^{42}$Sc and 
$^{56}$Ni $\to$ $^{58}$Cu in the 
$J^\pi=1^+$ [(a), (b)] and $J^\pi =0^+$ [(c), (d)] states smeared with a width of 0.1 MeV. 
For the $(J,T)=(1,0)$ channel, shown are the strengths obtained with factors $f=0, 1.0, 1.3, 1.7$, and 2.0. 
For the $(J,T)=(0,1)$ channel, the unperturbed single-particle transition strengths are also shown by a dotted line.}
\label{strength}
\end{center}
\end{figure}

Figure~\ref{strength} shows the strength distributions for the monopole ($L=0$) $pn$-pair-addition transfer 
$|\langle Z+1, N+1; \lambda | \hat{P}^\dagger_{T,S}| Z,N \rangle |^2 \equiv | \sum_{\alpha \beta} M^{T,S}_{\alpha \beta}|^2$ 
as functions of the RPA frequency $\omega_\lambda$ in $^{40}$Ca and $^{56}$Ni. 
Here, the $L=0$ $T=0$ $pn$-pair-addition operators are defined as
\begin{align}
\hat{P}^\dagger_{T=0, S=1, S_z} =& \notag \\ 
\dfrac{1}{2}\sum_{\sigma \sigma^\prime}\sum_{\tau}\int d\boldsymbol{r} &
\hat{\psi}^\dagger(\boldsymbol{r}\sigma \tau)\langle \sigma | \boldsymbol{\sigma}_{S_z}| \sigma^\prime \rangle
\hat{\psi}^\dagger(\boldsymbol{r}\tilde{\sigma}^\prime \tilde{\tau})  \label{pair1} 
\end{align}
and the $L=0$ $T=1$ $pn$-pair-addition operator as
\begin{align}
\hat{P}^\dagger_{T=1, T_z=0, S=0} = & \notag \\ 
\dfrac{1}{2}\sum_{\sigma}\sum_{\tau \tau^\prime}\int d\boldsymbol{r} &
\hat{\psi}^\dagger(\boldsymbol{r}\sigma \tau)\langle \tau | \boldsymbol{\tau}_0| \tau^\prime \rangle
\hat{\psi}^\dagger(\boldsymbol{r}\tilde{\sigma} \tilde{\tau}^\prime) \label{pair3}
\end{align}
in terms of the nucleon field operator, where 
$\hat{\psi}^\dagger(\boldsymbol{r} \tilde{\sigma} \tilde{\tau}) 
\equiv (-2\sigma)(-2\tau)\hat{\psi}^\dagger(\boldsymbol{r} -\sigma -\tau)$. 
Note that the absolute values of the RPA frequency do not directly correspond to the 
excitation energies observed experimentally. 
The particle excitation energies here are measured from the Fermi energies; 
$E_\alpha = |\epsilon_\alpha - \lambda|$.  
Since in the spatially spherical ``normal" nuclei, the spin orientation is not uniquely determined, 
i.e., rotationally invariant in spin space, 
the strengths for the spin-triplet $(S=1)$ pair-addition transfer (\ref{pair1}) are all the same. 
Therefore, the strengths for $S_z=0, \pm 1$ are summed up in Figs.~\ref{strength}(a) and  \ref{strength}(b). 

\begin{table}[t]
\begin{center}
\caption{Microscopic structure of the collective $J^\pi=1^+$ and $0^+$ states in $^{42}$Sc 
calculated with $f=1.7 (1.3)$. Listed are the configuration, 
its excitation energy, and the matrix element. 
The excitation energies are given in MeV. 
The pp and hh excitations possessing a large matrix element are only shown. 
Sums of the backward-going amplitudes squared and the matrix elements 
are shown in the last lines. For the $J^\pi=1^+$ state, the $J_z=0$ component is only shown.  }
\label{40Ca_amplitude}
\begin{tabular}{cccc}
\hline \hline
\noalign{\smallskip}
$^{42}$Sc & & $J^\pi=1^+$ & $J^\pi=0^+$   \\
\noalign{\smallskip}\hline\noalign{\smallskip}
configuration & $E_{\alpha}+E_{\beta}$ & $M^{S=1, S_z=0}_{\alpha \beta}$ & $M^{S=0}_{\alpha \beta}$ \\
\noalign{\smallskip}\hline\noalign{\smallskip}
$\pi 1f_{7/2} \otimes \nu 1f_{7/2} $ & 7.5 & 1.70 (0.92) & 2.82 \\  
$\pi 1f_{7/2} \otimes \nu 1f_{5/2}$ & 15.2 & 0.62 (0.38) & \\
$\pi 1f_{5/2} \otimes \nu 1f_{7/2}$ & 14.7 & 0.51 (0.31) & \\
$\pi 2p_{3/2} \otimes \nu 2p_{3/2}$ & 16.1 & 0.17 (0.11) & 0.15  \\
$\pi 1d_{3/2} \otimes \nu 1d_{3/2}$ & 4.2 & 0.16 (0.08) & 0.26 \\
$\pi 2s_{1/2} \otimes \nu 2s_{1/2}$ & 6.6 & 0.25 (0.12) & 0.09 \\
$\pi 1d_{3/2} \otimes \nu 1d_{5/2}$ & 10.1 & 0.32 (0.15) & \\
$\pi 1d_{5/2} \otimes \nu 1d_{3/2}$ & 10.2 & 0.32 (0.15) & \\
$\pi 1d_{5/2} \otimes \nu 1d_{5/2}$ & 16.1 & 0.16 (0.08) & 0.18 \\
\noalign{\smallskip}\hline\noalign{\smallskip}
 \multicolumn{2}{c}{$\sum_{\alpha \beta}M_{\alpha \beta}$} &  6.63 (4.51) & 4.56 \\
 \multicolumn{2}{c}{$\sum_{ij}Y_{ij}^2$}  &  0.17 (0.04)  & 0.03  \\
\noalign{\smallskip}
\hline \hline
\end{tabular}
\end{center}
\end{table}

One sees that the excitation energy and the strength of the $J^\pi = 1^+$ state 
are strongly affected by the $T=0$ pairing interaction. 
In the case of $f=0$, without the $T=0$ pairing interaction, the lowest $1^+$ state in $^{42}$Sc 
located at $\omega= 7.5$ MeV is a single-particle excitation $\pi f_{7/2}\otimes \nu f_{7/2}$. 
As the pairing interaction is switched on and the strength is increased, 
the $1^+$ state is shifted lower in energy with the enhancement of the transition strength. 
In Table~\ref{40Ca_amplitude}, the microscopic structure of the $1^+$ state obtained 
by setting $f$ to 1.7 and 1.3 (in parentheses) is summarized. 
This $1^+$ state is constructed by many pp excitations involving 
an $f_{5/2}$ and a $p_{3/2}$ orbitals located above the Fermi levels as well as the 
$\pi f_{7/2}\otimes \nu f_{7/2}$ excitation. 
It is particularly worth noting that the hh excitations from the $sd$-shell have an appreciable contribution 
to generate this $T=0$ $pn$-pair-addition vibrational mode, 
indicating a $^{40}$Ca core-breaking. 
Furthermore, all the pp and hh excitations listed in the table construct the vibrational mode in phase. 
The strong collectivity can be also seen from a large amount of the ground-state correlation: 
A sum of the backward-going amplitudes squared is 0.17 (0.04).

\begin{table}[t]
\begin{center}
\caption{Same as Table~\ref{40Ca_amplitude} but for $^{58}$Cu.}
\label{56Ni_amplitude}
\begin{tabular}{cccc}
\hline \hline
\noalign{\smallskip}
$^{58}$Cu & & $J^\pi=1^+$ & $J^\pi=0^+$   \\
\noalign{\smallskip}\hline\noalign{\smallskip}
configuration & $E_{\alpha}+E_{\beta}$ & $M^{S=1, S_z=0}_{\alpha \beta}$ & $M^{S=0}_{\alpha \beta}$ \\
\noalign{\smallskip}\hline\noalign{\smallskip}
$\pi 2p_{3/2} \otimes \nu 2p_{3/2}$ & 4.5 & 1.28 (1.38) & 1.95 \\
$\pi 2p_{1/2} \otimes \nu 2p_{3/2}$ & 6.4 & 0.39 (0.28) & \\
$\pi 2p_{3/2} \otimes \nu 2p_{1/2}$ & 6.5 & 0.37 (0.26) & \\
$\pi 2p_{1/2} \otimes \nu 2p_{1/2}$ & 7.9 & 0.05 (0.03) &  0.16 \\
$\pi 1f_{5/2} \otimes  \nu 1f_{5/2}$ & 9.7 & 0.15 (0.09) & 0.33  \\
$\pi 1g_{9/2} \otimes \nu 1g_{9/2}$ &  17.7 & 0.24 (0.14) & 0.16 \\ 
$\pi 1f_{7/2} \otimes \nu 1f_{7/2}$ & 5.1 & 0.17 (0.13) & 0.07 \\ 
\noalign{\smallskip}\hline\noalign{\smallskip}
 \multicolumn{2}{c}{$\sum_{\alpha \beta}M_{\alpha \beta}$} &  4.25 (3.19) & 3.53 \\
 \multicolumn{2}{c}{$\sum_{ij}Y_{ij}^2$} &  0.03 (0.008) & 0.009 \\
\noalign{\smallskip}
\hline \hline
\end{tabular}
\end{center}
\end{table}

The low-lying $1^+$ state in $^{58}$Cu is also sensitive to the $T=0$ pairing interaction. 
As shown in Table~\ref{56Ni_amplitude}, this mode 
 is dominantly constructed by a $\pi p_{3/2}\otimes \nu p_{3/2}$ 
excitation together with many other pp excitations involving a $p_{1/2}$ and an $f_{5/2}$ orbitals. 
In contrast to a large core breaking in $^{42}$Sc, a role played by the hh excitation of $\pi f_{7/2}\otimes \nu f_{7/2}$ 
is minor in $^{58}$Cu with the same pairing interaction.

In Figs.~\ref{strength}(c) and \ref{strength}(d), 
the strength distributions for 
the $T=1$ $pn$-pair-addition transfer 
are shown together with the strengths obtained without the residual interactions.
The low-lying $0^+$ state is predominantly constructed by the $\pi f_{7/2}\otimes \nu f_{7/2}$ excitation in $^{42}$Sc 
similarly to the $1^+$ state. 
Though the number of possible pp configuration in the bound states 
is smaller than in the $T=0$ channel, the energy shift due to 
the $T=1$ pairing interaction is large and the ground-state correlation is strong. 
The $0^+$ state in $^{58}$Cu is as collective as the $1^+$ state.

\begin{figure}[t]
\begin{center}
\includegraphics[scale=0.26]{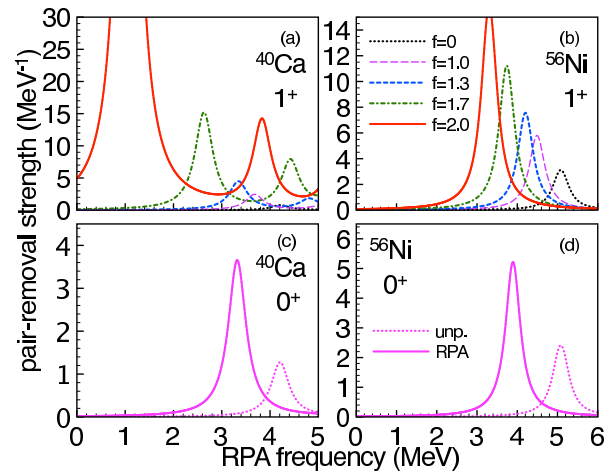}
\caption{(Color online) Same as Fig.~\ref{strength} but for the $pn$ pair-removal strengths.}
\label{strength_rem}
\end{center}
\end{figure}

In an attempt to explore characteristic features of the collective $T=0$ $pn$-pairing vibration, 
I investigate the $pn$ pair-removal strengths in $^{40}$Ca and $^{56}$Ni.  
The strength distributions for the $pn$-pair removal transfer 
are shown in Fig.~\ref{strength_rem}.  
Similarly to the $T=0$ $pn$-pair-addition vibration, 
the frequency and the transition strength 
to the low-lying $1^+$ state 
strongly depend on the strength of the $T=0$ pairing interaction, 
in particular, for $^{40}$Ca $\to$ $^{38}$K. 
In the case of $f=1.7$, the $1^+$ state is mainly generated by a $\pi d_{3/2}\otimes \nu d_{3/2}$ excitation with 
a matrix element of 0.82. Furthermore, many other hh excitation participate to generate this $T=0$ $pn$-pair-removal vibrational mode: 
they are the $\pi s_{1/2}\otimes \nu s_{1/2}$ (with $M=0.07$), $\pi d_{5/2}\otimes \nu d_{3/2}$ (0.30), 
$\pi d_{3/2}\otimes \nu d_{5/2}$ (0.30), $\pi d_{5/2}\otimes \nu d_{5/2}$ (0.16) excitations 
together with the pp excitation of $\pi f_{7/2} \otimes \nu f_{7/2}$ (0.39). 
One sees the coherence among the hh and pp excitations, and 
a strong ground-state correlation: $\sum_{mn}Y_{mn}^2 = 0.11$.

\begin{figure}[t]
\begin{center}
\includegraphics[scale=0.21]{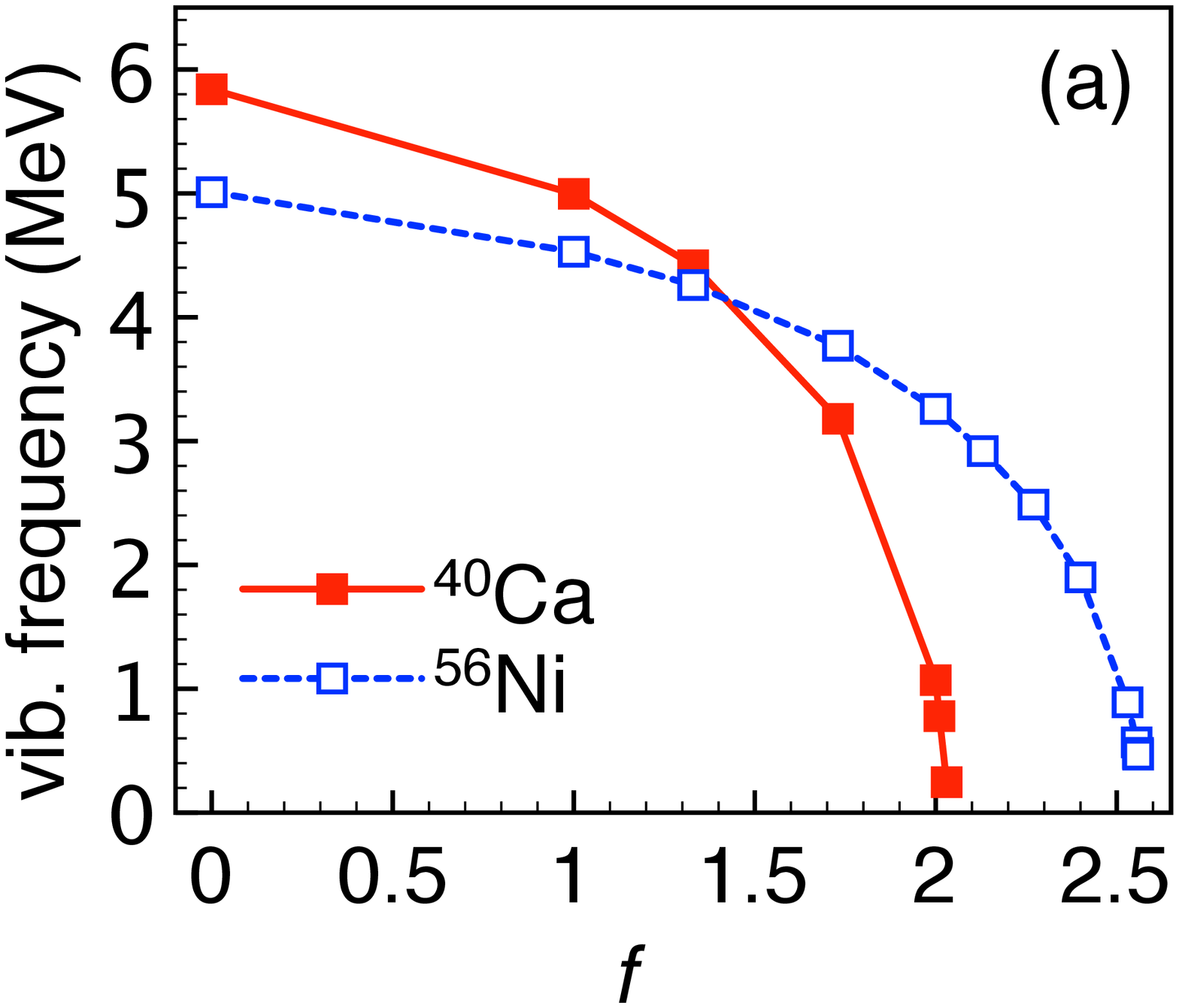}
\includegraphics[scale=0.21]{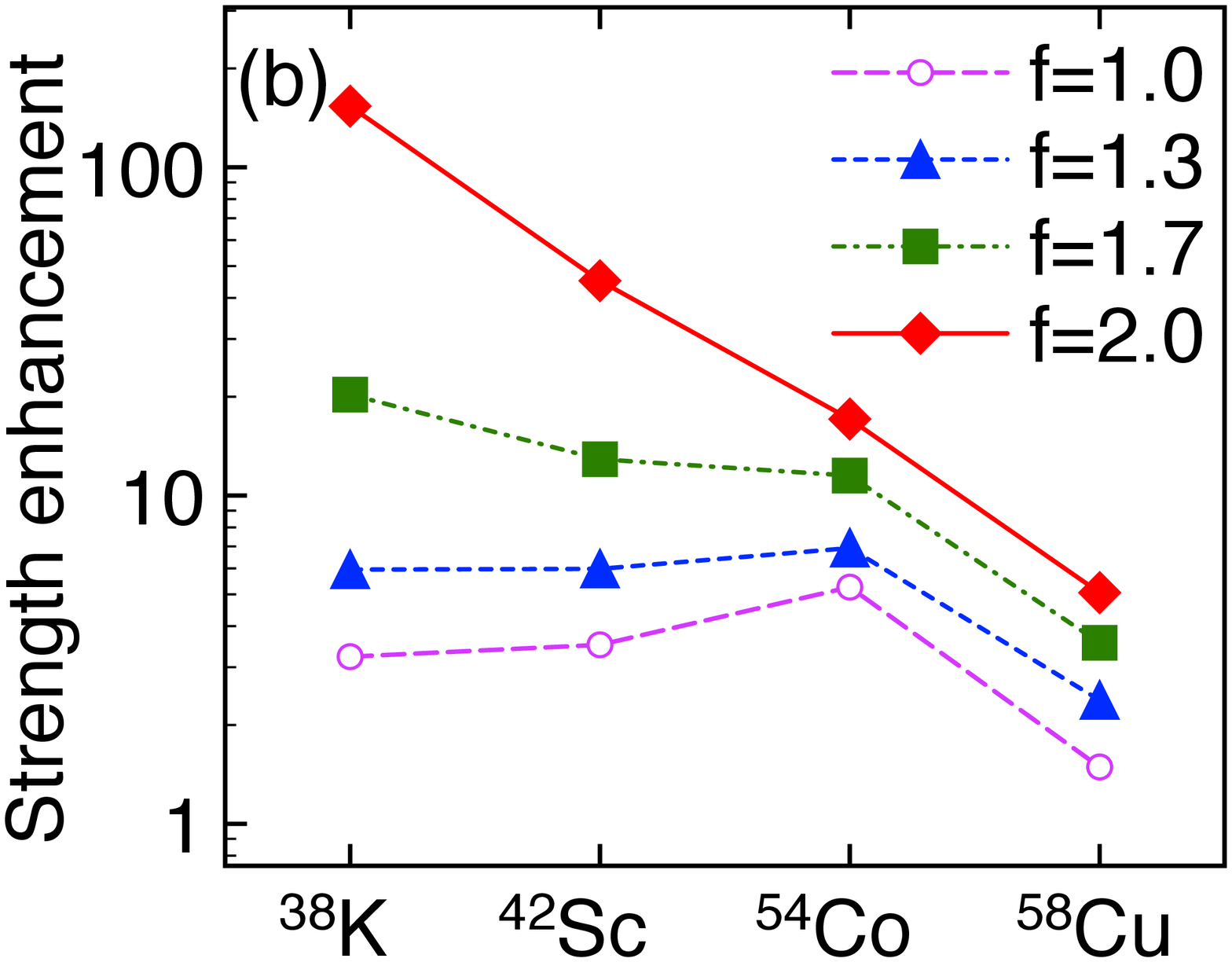}
\caption{(Color online) 
(a) RPA frequency of $T=0$ pairing vibrational mode in $^{40}$Ca and $^{56}$Ni 
calculated by varying the strength $f$. 
(b) Ratio of the transition strengths to the collective $1^+$ state 
to the ones to the unperturbed $1^{+}$ state 
in $^{38}$K, $^{42}$Sc, $^{56}$Co, and $^{58}$Cu calculated varying the pairing strength $f$. 
Lines are drawn to guide the eye.  
}
\label{freq}
\end{center}
\end{figure}

A change in the RPA frequency of the collective mode 
due to the $T=0$ pairing interaction is summarized in Fig.~\ref{freq}(a). 
The vibrational frequency is defined as $(\omega^{T=0}_{\mathrm{add}} + \omega^{T=0}_{\mathrm{rem}})/2$, 
where $\omega^{T=0}_{\mathrm{add (rem)}}$ 
denotes the RPA frequency of the eigenmode possessing the largest $pn$ pair-addition (removal) strength 
in the low energy region less than 10 MeV. 
In the doubly-magic nuclei, the pairing collectivity is generated by only the residual pairing interactions (\ref{v_pp_0}) and (\ref{v_pp_1}). 
One clearly sees that the 
RPA frequency of the $T=0$ $pn$-pairing vibrational mode becomes lower with increasing the pairing strength $f$. 
The pairing collectivity generated is sensitive to the shell structure as well as the interactions. 
The critical strength is $f_{\mathrm{c}} = 2.04$ and $2.57$ in $^{40}$Ca and $^{56}$Ni, respectively. 
It is known that the RPA breaks down at the critical point and underestimates the excitation energy
around that point~\cite{RS}. 
A rapid lowering of the RPA frequency 
seen here indicates that we have a true vacuum 
giving the $T=0$ pairing gaps  $\boldsymbol{\Delta}\equiv \langle \hat{P}_{T=0,S=1} \rangle \ne 0$ 
in the limit of the strong pairing interaction $f > f_{\mathrm{c}}$. 
Therefore the $1^+$ state can be considered as a  precursory soft mode of the $T=0$ pairing condensation.

Another direct measure of the collectivity is the $pn$ transfer strength. 
Figure~\ref{freq}(b) shows 
the ratio of the transition strengths to the collective $1^+$ state 
to the ones to the unperturbed $1^{+}$ state assuming the single-particle configuration with the lowest energy 
in $^{38}$K, $^{42}$Sc, $^{56}$Co, and $^{58}$Cu, that is a 
$(d_{3/2})^{-2}, (f_{7/2})^2, (f_{7/2})^{-2}$, and $(p_{3/2})^2$ configuration, respectively. 
In $^{38}$K and $^{42}$Sc, we see an exponential enhancement of the transition strengths 
when approaching the critical strength $f_\mathrm{c}$.  
It is noted that the deuteron transfer experiment was performed by using the 
$^{40}$Ca($^3$He, p)$^{42}$Sc reaction, and 
the observed cross section to the lowest $1^+$ state is about 24 times as large as 
the cross section calculated assuming the pure $(f_{7/2})^2$ configuration~\cite{puh68}.  
It is thus interesting to see in a future work the $pn$-transfer cross sections calculated 
by using the microscopic transition densities in the present framework.

I am going to discuss here how the strength $f$ is fixed. 
An analysis made in Ref.~\cite{ber10} suggests $f \simeq 1.6$ 
for the density-independent contact interactions 
based on the phenomenological shell-model Hamiltonians 
in the $fp$-shell nuclei.
The pairing strengths can be also determined from the proton-neutron scattering lengths in 
the $T=0$ and $1$ channels~\cite{esb97} 
as $f \sim 1.4$ for $E_\mathrm{cut}=60$ MeV, 
and the low-lying states in $N=Z$ odd-odd nuclei were investigated 
by employing the density-dependent pairing interaction thus determined~\cite{tan14}. 
The authors in Ref.~\cite{bai13} pointed out that the low-lying Gamow-Teller (GT) strengths in $N \simeq Z$ nuclei 
are sensitive to the $T=0$ pairing interaction. 
Thus, the low-lying GT strengths in the neighboring nuclei can be alternatively used to 
fix the $f$ value.

Quite recently, enhancement of the GT strengths 
to the low-energy region in the $N=Z$ odd-odd nuclei in the $fp$-shell
was reported and the low-lying strengths are found to be 
very sensitive to the $T=0$ pairing interaction~\cite{fuj14}. 
The GT strengths of $^{42}$Ca to the low-lying $1^+$ state in $^{42}$Sc are particularly enhanced. 
Without the residual interactions, the low-lying and high-lying GT modes correspond primarily to 
the $\pi f_{7/2} \otimes \nu f_{7/2}$ and $\pi f_{5/2} \otimes \nu f_{7/2}$ excitations, respectively. 
The enhanced strength to the low-lying states indicates a coherent contribution of these excitations. 
As shown in Table~\ref{40Ca_amplitude}, the lowest $1^+$ state in $^{42}$Sc is generated by the 
 $\pi f_{7/2} \otimes \nu f_{7/2}$ excitation together 
with the high-lying $\pi f_{5/2} \otimes \nu f_{7/2}$ excitation due to the $T=0$ pairing interaction. 
The result reported in Ref.~\cite{fuj14} stimulates a further investigation 
on how the $pn$-pairing collectivity of the low-lying $1^+$ state 
is seen in the GT strength of a ph-type, 
while the GT strengths associated with the $pn$-pairing collectivity were investigated 
in a solvable model~\cite{eng97}, 
and in the context of the SU(4) symmetry in the spin-isospin space within a Skyrme EDF framework~\cite{bai13}. 

\begin{figure}[t]
\begin{center}
\includegraphics[scale=0.22]{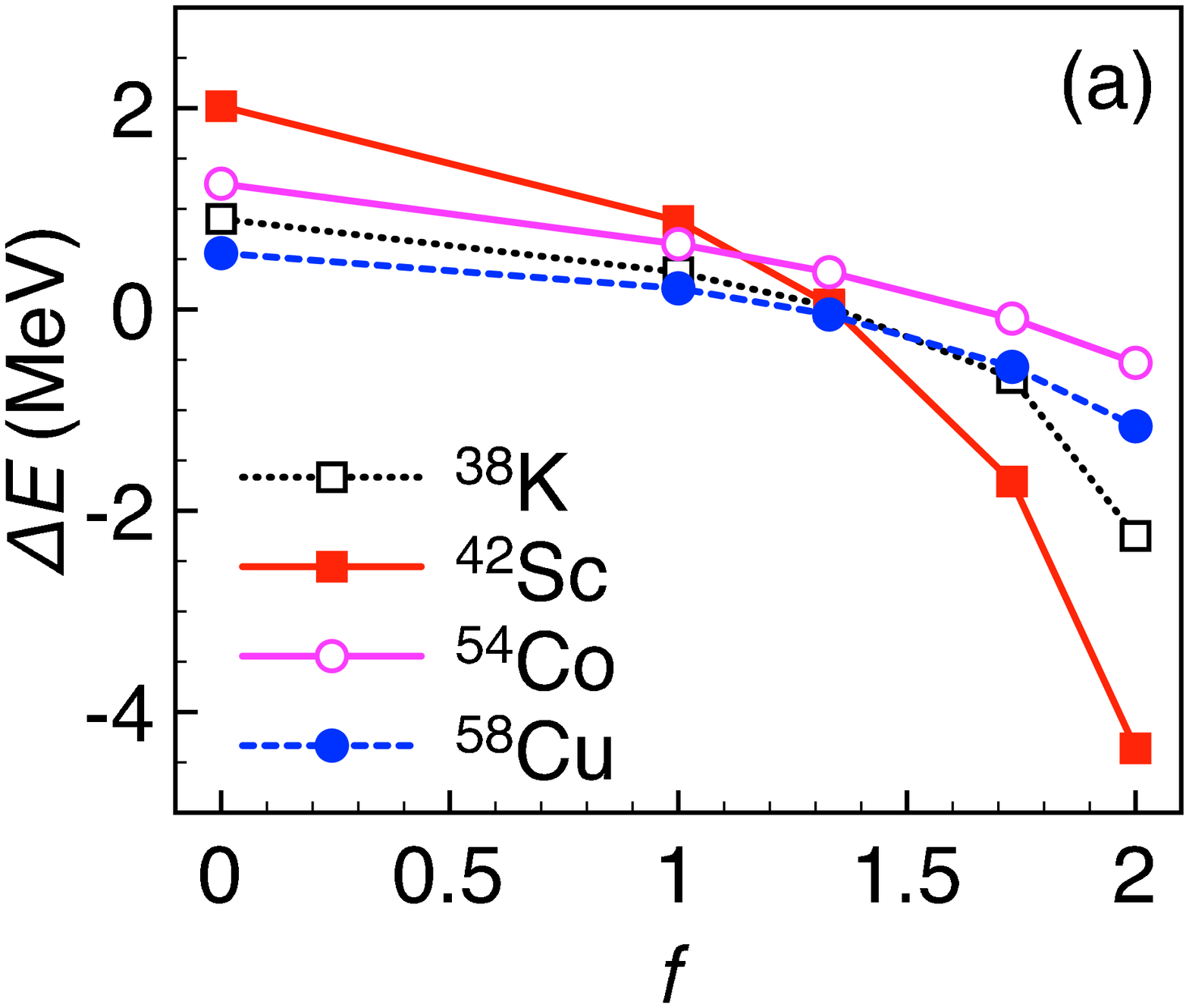}
\includegraphics[scale=0.22]{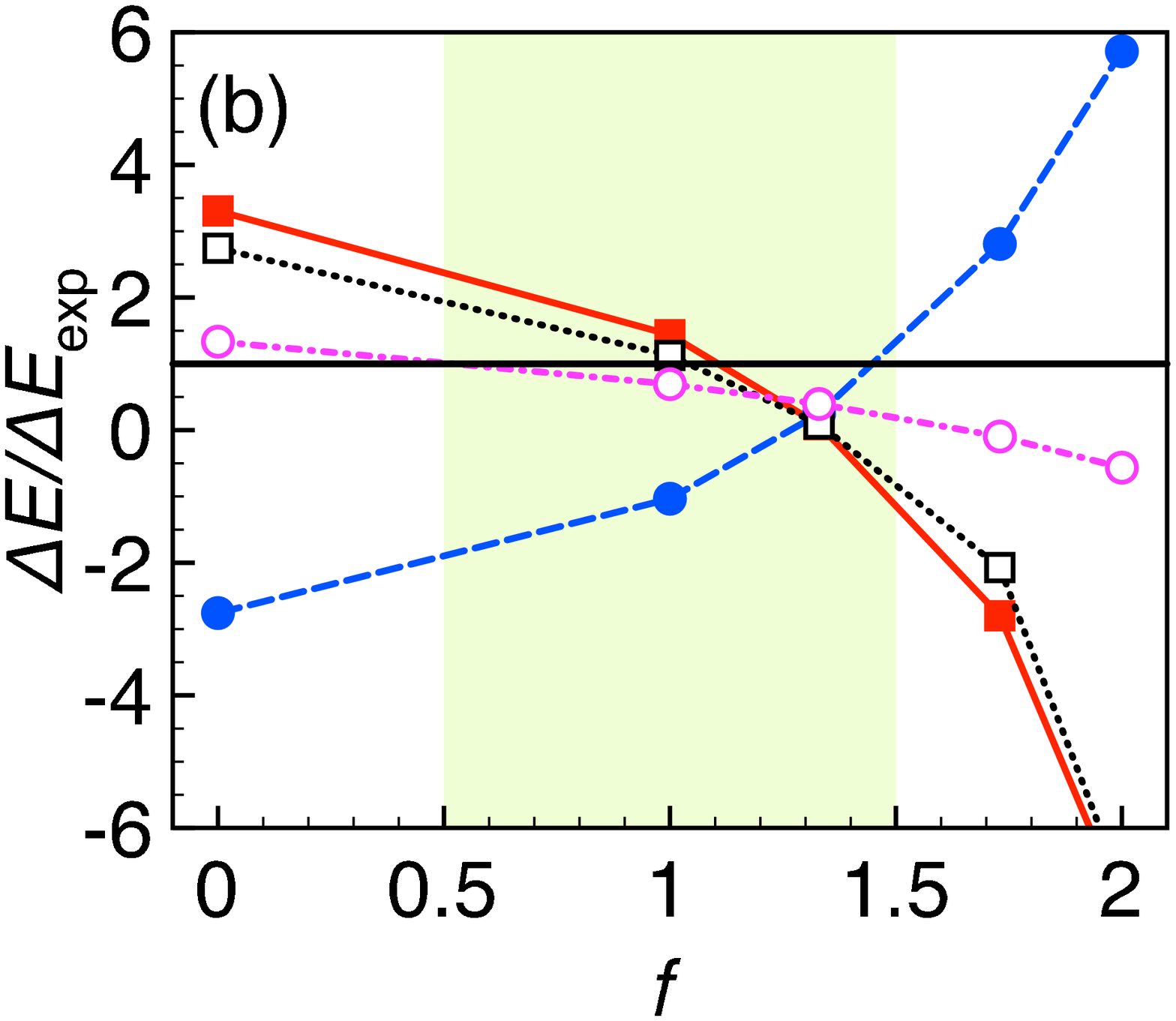}
\caption{(Color online) (a) Energy difference $\Delta E = \omega_{1^+}-\omega_{0^+}$ in 
$^{38}$K, $^{42}$Sc, $^{54}$Co and $^{58}$Cu 
calculated with $f=0$, 1.0, 1.3, 1.7, and 2.0. 
(b) Ratio of the energy difference calculated to the experimental value $\Delta E/ \Delta E_{\mathrm{exp}}$. 
The experimental data are taken from Ref.~\cite{nndc}. 
A horizontal line represents unity.
Lines are drawn to guide the eye.   
}
\label{delta_e}
\end{center}
\end{figure}

An energy difference between the $1^+$ and $0^+$ states is plotted in Fig.~\ref{delta_e}(a) 
and shown in Fig.~\ref{delta_e}(b) are the ratios of the energy difference $\Delta E$ 
calculated varying the strength $f$ 
to the one  experimentally observed:  
$\Delta E_{\mathrm{exp}} = E_{1^+_1}-E_{0^+_1}$ is $0.328, 0.611, 0.937$,  and $-0.203$ MeV 
in $^{38}$K, $^{42}$Sc, $^{54}$Co, and $^{58}$Cu, respectively~\cite{nndc}. 
The pairing strength $V_0$ was fixed to reproduce the experimental value for the 
$T=1$ pairing vibrational frequency, 4.39 and 4.07 MeV in $^{40}$Ca and $^{56}$Ni. 
It is defined by the binding energies; $B(Z,N)-[B(Z+1,N+1) - B(Z-1, N-1)]/2$. 
The strength $V_0 = -390$ MeV fm$^3$ produces 4.38 and 4.10 MeV in $^{40}$Ca and $^{56}$Ni, respectively 
for the vibrational frequency defined as $(\omega^{T=1}_{\mathrm{add}} + \omega^{T=1}_{\mathrm{rem}})/2$. 
The calculated results obtained by using $f=0.5 - 1.5$ [a shaded area in Fig.~\ref{delta_e}(b)] 
reproduce the experimental data for the energy difference. 
However, one sees that we have a large uncertainty for determining the strength $f$. 
Since the pairing interaction in the $T=0$ channel is crucial for 
a quantitative discussion on the collectivity of the pairing vibrational modes, 
it is largely desired to investigate it in detail, 
such as the density dependence of the interaction, 
and the mass number and/or the isospin dependence of the strength as introduced in Ref.~\cite{niu13}. 

Before summarizing the paper, it is noted concerning the $T=1$ pairing that 
the pairing strength $V_0 = -390$ MeV fm$^3$ gives $\Delta_\nu = 0.97$ MeV and $\Delta_\pi = 0.99$ MeV in $^{44}$Ti 
by solving the SHF-Bogoliubov equation with an energy cut off at 60 MeV 
and assuming that the $T=1$ pairing interaction is rotationally invariant in isospace. 
The experimental pairing gaps of neutrons and protons are 2.06 and 1.86 MeV. 
Thus, the resultant pairing correlation in the ground state is very weak. 
In the present framework of the HF+RPA employing 
the Skyrme SGII and the density-dependent pairing EDFs, 
we cannot describe consistently the static and dynamic $T=1$ pairing correlations in a unified way.

To summarize, I have found that a collective $T=0$ $pn$-pairing vibrational mode emerges 
in the presence of the $T=0$ two-body particle-particle interaction 
in a self-consistent Skyrme-EDF framework. 
It is suggested that the low-lying $J^{\pi} =1^+$ state in odd-odd $N = Z$ nuclei can be 
a precursory soft mode of the $T=0$ pairing condensation. 
Due to a strong collectivity of the $T=0$ $pn$-pairing vibration,
the $pn$-transfer strength to the $1^+$ state can be largely enhanced in comparison with 
the strength made by the pure single-particle configuration. 
The present framework, however, cannot account for all the experimental information 
on the pairing correlations in a consistent way. 
For a quantitative discussion, it is greatly desirable to investigate 
the Skyrme and pairing EDFs 
both in the $T=0$ and $T=1$ channels in more details.

Valuable discussions with M.~Matsuo, K.~Matsuyanagi, and H.~Sagawa are acknowledged. 
This work was supported by KAKENHI Grant Nos. 23740223 and 25287065. 
The numerical calculations were performed on SR16000
at the Yukawa Institute for Theoretical Physics, Kyoto University and 
on T2K-Tsukuba and COMA, at the Center for Computational Sciences, University of Tsukuba. 
Some of the results were obtained by using the K computer at the RIKEN 
Advanced Institute for Computational Science (hp120192, hp140001).

\end{document}